\documentclass{article}

\usepackage{cite}
\usepackage{graphicx}
\usepackage{dcolumn}

\begin{document}

\title{Comment on ``Bidimensional bound states for charged polar nanoparticles''}
\author{Paolo Amore \\
%EndAName
Facultad de Ciencias, CUICBAS, Universidad de Colima, \\
Bernal D\'{\i}az del Castillo 340, Colima, Colima, Mexico \thanks{%
paolo.amore@gmail.com}\\
Francisco M. Fern\'andez \\
INIFTA (CONICET), Divisi\'{o}n Qu\'{i}mica Te\'{o}rica, \\
Blvd. 113 y 64 (S/N), Sucursal 4, Casilla de Correo 16, \\
1900 La Plata, Argentina \thanks{%
framfer@gmail.com}}
\maketitle

\begin{abstract}
In a recent paper Castellanos-Jaramillo and Castellanos-Moreno proposed a
simple quantum-mechanical model for an electron in the vicinity of an
ionized nanostructure with a permanent electric dipole. They chose the
interaction of the electron with the charge and the dipole in such a way
that the resulting Schr\"{o}dinger equation is separable into radial and
angular parts. In this comment we show that those authors did not solve the
angular eigenvalue equation with proper periodic boundary conditions and
that they also made a mistake in the elimination of the first derivative in
the radial equation. Such errors invalidate their results of the Einstein
coefficients for the $\left( GaAs\right) _{3}$ system considered.
\end{abstract}

\section{Introduction}

\label{sec:intro}

In a recent paper Castellanos-Jaramillo and Castellanos-Moreno\cite{CC19}
(CC from now on) put forward a simple quantum-mechanical model for an
electron in the vicinity of an ionized nanostructure with a permanent
electric dipole. They chose the interaction of the electron with the charge
and the dipole in such a way that the resulting Schr\"{o}dinger equation is
separable into radial and angular parts. At first they state that the
solutions to the latter eigenvalue equation should be periodic of period $%
2\pi $ but later they turn to somewhat different boundary conditions.

The purpose of this paper is to solve the Schr\"{o}dinger equation for the
model proposed by those authors with true periodic boundary conditions in
order to determine to which extent this change may affect their results. In
section~\ref{sec:model} we transform the Schr\"{o}dinger equation into a
dimensionless eigenvalue equation and calculate the eigenvalues of the
angular part when the eigenfunctions a periodic functions of period $2\pi $.
We compare present results with those obtained by CC. Finally, in section~%
\ref{sec:model} we discuss the results and draw conclusions.

\section{The model}

\label{sec:model}

The model Hamiltonian chosen by CC\cite{CC19} is
\begin{equation}
H=-\frac{\hbar ^{2}}{2m}\nabla ^{2}+\frac{qQ}{4\pi \epsilon _{0}\left|
\mathbf{r}-\mathbf{r}_{0}\right| }+\frac{qD\cos \theta }{4\pi \epsilon
_{0}\left| \mathbf{r}-\mathbf{r}_{0}\right| ^{2}}+Be^{-r^{2}/\sigma ^{2}}
\end{equation}
where $\mathbf{r}=(x,y)$, $q=-e$ and $Q=Ze$ are the charges of the electron
and the nanostructure, respectively, $D$ is the dipole of the latter, $%
\epsilon _{0}$ is the vacuum permittivity and $\sigma $ is related
to the
radius of the nanoparticle (note that we write $\sigma ^{2}$ instead of $%
\sigma $ in the Gaussian term).

In order to facilitate the treatment of the Schr\"{o}dinger
equation it is convenient to transform it into a dimensionless
eigenvalue equation. CC do it in a rather confuse way; therefore,
we proceed differently. By means of the change of variables
\begin{equation}
\mathbf{r}=L\mathbf{\rho },\;L=-\frac{4\pi \epsilon _{0}\hbar ^{2}}{mqQ}
\end{equation}
the Schr\"{o}dinger equation $H\psi =E\psi $ becomes
\begin{eqnarray}
H^{\prime }\Phi &=&\mathcal{E}\Phi  \nonumber \\
H^{\prime } &=&-\frac{1}{2}\nabla ^{\prime 2}-\frac{1}{\left| \mathbf{\rho }-%
\mathbf{\rho }_{0}\right| }+\frac{\xi \cos \theta }{\left| \mathbf{\rho }-%
\mathbf{\rho }_{0}\right| ^{2}}+Ae^{-a^{2}\rho ^{2}}  \nonumber \\
\xi &=&\frac{mqD}{4\pi \epsilon _{0}\hbar ^{2}},\;A=\frac{mL^{2}B}{\hbar ^{2}%
},\;a=\frac{L}{\sigma },\;\mathbf{\rho }_{0}=\frac{\mathbf{r}_{0}}{L}
\nonumber \\
\mathcal{E} &=&\frac{mL^{2}E}{\hbar ^{2}}
\end{eqnarray}

Following CC we choose $\mathbf{r}_{0}=0$ so that the equation is separable
in polar coordinates
\begin{equation}
H=-\frac{1}{2\rho }\frac{\partial }{\partial \rho }\rho \frac{\partial }{%
\partial \rho }-\frac{1}{2\rho ^{2}}\frac{\partial ^{2}}{\partial \theta ^{2}%
}-\frac{1}{\rho }+\frac{\xi \cos \theta }{\rho ^{2}}+Ae^{-a^{2}\rho ^{2}}
\label{eq:H_spherical}
\end{equation}
Upon setting $\Phi (\rho ,\theta )=f(\rho )g(\theta )$ and choosing $%
g(\theta )$ so that
\begin{equation}
\left( -\frac{1}{2}\frac{d^{2}}{d\theta ^{2}}+\xi \cos \theta \right)
g=\lambda g  \label{eq:eigen_rot}
\end{equation}
then
\begin{equation}
\left( -\frac{1}{2\rho }\frac{\partial }{\partial \rho }\rho \frac{\partial
}{\partial \rho }-\frac{1}{\rho }+\frac{\lambda }{\rho ^{2}}+Ae^{-a^{2}\rho
^{2}}\right) f=\mathcal{E}f  \label{eq:eigen_radial}
\end{equation}
CC first state that the solutions to equation (\ref{eq:eigen_rot}) should be
periodic of period $2\pi $ ($g(\theta +2\pi )=g(\theta )$) but later they
turn to different boundary conditions.

The eigenvalue equation (\ref{eq:eigen_rot}) with periodic boundary
conditions has discrete eigenvalues $\lambda =\lambda _{m}$, $m=0,1,\ldots $
($\lambda _{0}<\lambda _{1}<\ldots $) and those of (\ref{eq:eigen_radial})
will be $\mathcal{E}_{nm}$, $n=0,1,\ldots $. In order to compare present
results with those of CC\cite{CC19} note that $2\xi =gp$ and $2\lambda
=\lambda ^{CC}$.

For $\lambda \geq 0$ the behaviour of the solution to the radial equation (%
\ref{eq:eigen_radial}) at the origin is $f(\rho )\approx \rho ^{-\sqrt{%
2\lambda }}$. On the other hand, when $\lambda <0$ this function goes
through infinitely many zeroes when $\rho \rightarrow 0$ and the spectrum
becomes continuous and unbounded from below. In order to overcome this
difficulty one may define a self-adjoint extension of the Hamiltonian by
specifying a particular boundary condition at $\rho =0$\cite{CH02}. We will
not discuss this aspect of the problem in detail here because we will not
need to solve the radial equation. In order to be square integrable the
behaviour of the solution to the radial equation at infinity should be $%
f(\rho )\approx e^{-\sqrt{-2\mathcal{E}}\rho }$. Consequently, there are
simple suitable solutions $\Phi _{nm}(\rho ,\theta )$ for the bound-states
of the Schr\"{o}dinger equation when $\lambda \geq 0$ and $\mathcal{E}<0$
and some additional care is required for $\lambda <0$. CC\cite{CC19}
bypassed the problem of negative eigenvalues $\lambda _{m}$ by postulating
that they are physically unacceptable.

For large values of $\xi $ the eigenvalues of the angular equation behave as
\begin{equation}
\lambda _{m}\approx -\xi +\sqrt{\xi }\left( m+\frac{1}{2}\right) +\mathcal{O}%
(1)  \label{eq:lambda_m_asympt}
\end{equation}
Therefore, there are values $\xi =\xi _{m_{c}}$ such that $\lambda
_{m_{c}}=0 $ and, consequently, $\lambda _{m}<0$ for all $m<m_{c}$. In order
to obtain the critical values $\xi _{m_{c}}$ it is only necessary to solve
equation (\ref{eq:eigen_rot}) for $\lambda =0$.

\begin{figure}[tbp]
\begin{center}
\includegraphics[width=9cm]{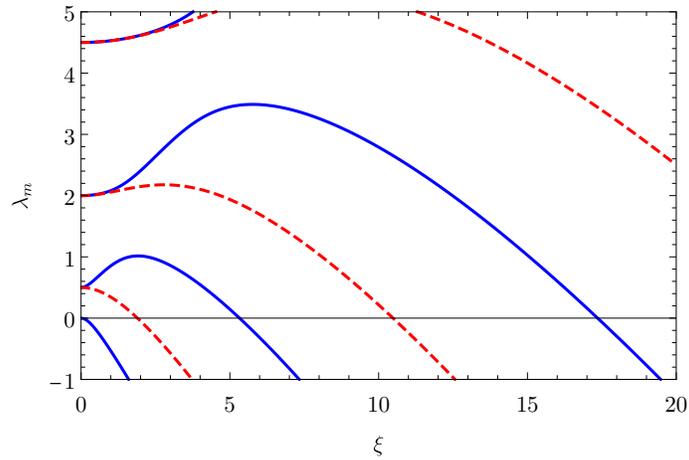}
\end{center}
\caption{Eigenvalues $\lambda_m$ of the angular equation for a range of $\xi$
values. The continuous (blue) and broken (red) curves denote even and odd
solutions, respectively}
\label{fig:ROTEN}
\end{figure}

The even and odd solutions to the angular equation (\ref{eq:eigen_rot}) can
be expanded in Fourier series of period $2\pi $%
\begin{eqnarray}
g^{e}(\theta ) &=&\sum_{j=0}^{\infty }a_{j}\cos (j\theta )  \nonumber \\
g^{o}(\theta ) &=&\sum_{j=1}^{\infty }b_{j}\sin (j\theta )
\label{eq:Fourier}
\end{eqnarray}
respectively. The coefficients $a_{j}$ and $b_{j}$ can be easily obtained as
polynomial functions of $\lambda $ from simple three-term recurrence
relations:
\begin{eqnarray}
&&2\lambda a_{0}-\xi a_{1} =0,\;\left( 2\lambda -1\right) a_{1}-\xi \left(
2a_{0}+a_{2}\right) =0  \nonumber \\
&&\left( 2\lambda -n^{2}\right) a_{n}-\xi \left( a_{n-1}+a_{n+1}\right) ,\;n
=2,3,\ldots  \nonumber \\
&&\left( 2\lambda -n^{2}\right) b_{n}-\xi \left( b_{n-1}+b_{n+1}\right) ,\;n
=1,2,\ldots ,\;b_{0}=0  \label{eq:rec_rel}
\end{eqnarray}
From the termination conditions $a_{N}(\lambda ,\xi )=0$ or $b_{N}(\lambda
,\xi )=0$ we obtain $\lambda (\xi )$ or $\xi (\lambda )$ with any desired
accuracy provided that $N$ is large enough. Setting $\lambda =0$ we obtain
the critical values $\xi _{m_{c}}$ mentioned above; the first of them are $%
\xi _{0}=0$, $\xi _{1}=1.894922593$, $\xi _{2}=5.324657803$. Note that $%
\lambda _{0}(\xi )$ is negative for all $\xi >0$. Fig.~\ref{fig:ROTEN} shows
the first eigenvalues $\lambda _{m}$ for a range of $\xi $ values.

CC\cite{CC19} chose a set of model parameters that appear to be suitable for
$\left( GaAs\right) _{3}$ and obtained $4gp=0.8147872$. Table~\ref
{tab:lambdas} shows that our results $\lambda ^{CC}=2\lambda $ for $\xi
=0.8147872/8$ do not agree with those in Table~1 of CC. The reason is
probably that CC did not use proper periodic angular eigenfunctions; note
that they claim to have used equation (33) ($y(z)=e^{\nu z}\phi (z)$)
instead of the correct one $y(z)=e^{i\nu z}\phi (z)$.

\begin{figure}[tbp]
\begin{center}
\includegraphics[width=6cm]{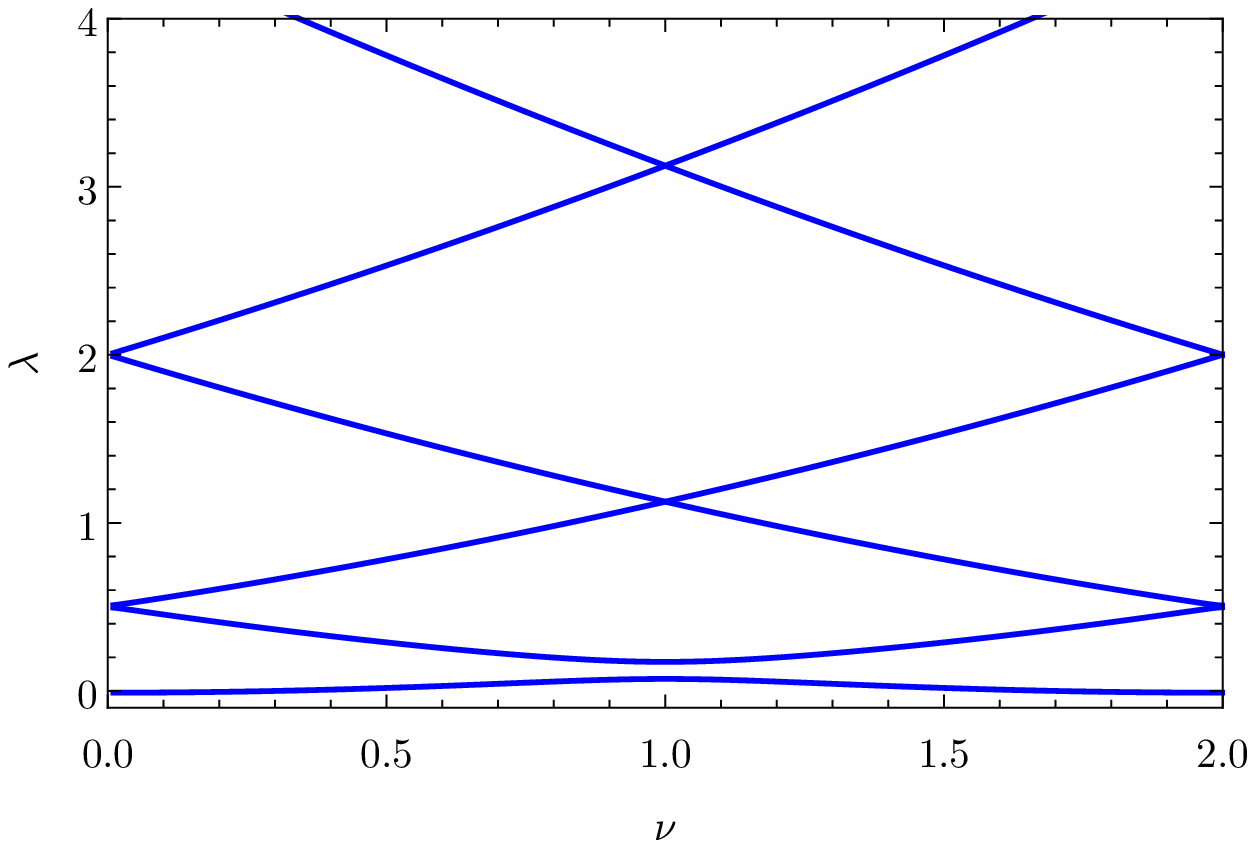} %
\includegraphics[width=6cm]{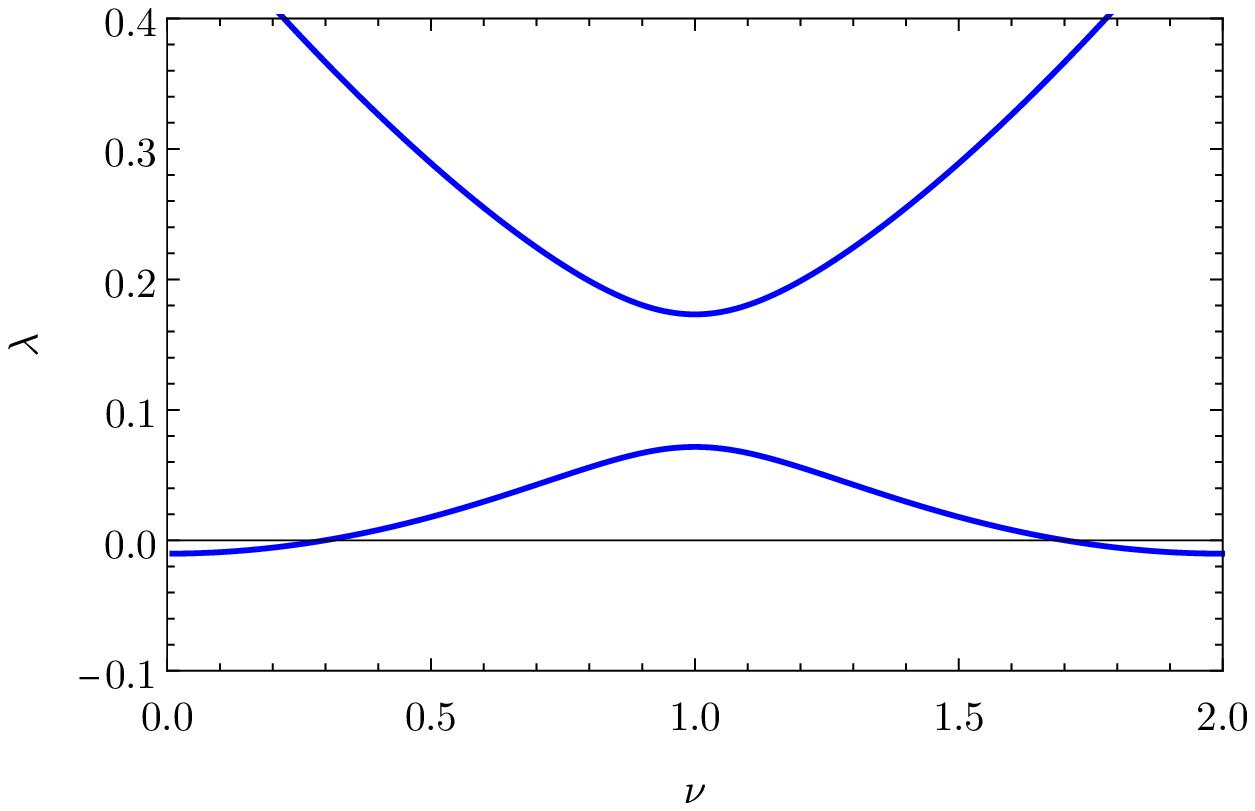}
\end{center}
\caption{First eigenvalues $\lambda_m(\nu)$ for $\xi=0.8147872/8$}
\label{fig:lambda_vs_nu}
\end{figure}

If we define $z=\theta /2$ then $u(z)=g(2z)$ is a periodic function of
period $\pi $ that we may rewrite as $u(z)=e^{i\nu z}v(z)$, where $v(z)$ is
periodic of period $\pi $. Note that $u(z)$ will be periodic of period $\pi $
provided that $\nu =0,2,\ldots $. If we solve the eigenvalue equation for $%
v(z)$ by means of its expansion in the basis set $\phi _{j}=\frac{1}{\sqrt{%
\pi }}e^{2ijz}$, $j=0\pm 1,\pm 2,\ldots $ we obtain the eigenvalues $\lambda
_{m}(\nu )$ shown in figure~\ref{fig:lambda_vs_nu}. Note that $\lambda
_{m}(\nu =0)=\lambda _{m}(\nu =2)$ as expected and the interesting fact that
the eigenvalues exhibit avoided crossings at $\nu =1$. The left panel shows
the first $6$ eigenvalues. If we just consider the first two ones (right
panel) then we realize that the discontinuity in CC's figure~1 may probably
come from choosing the lowest eigenvalue for $0<\nu <1$ and the first
excited one for $\nu >1$.

We now briefly turn to the radial equation. If we write $f(\rho )=u(\rho )/%
\sqrt{\rho }$ we obtain
\begin{equation}
u^{\prime \prime }(\rho )+\left[ 2\mathcal{E}+\frac{2}{\rho }-2Ae^{a^{2}\rho
^{2}}-\frac{2\lambda -\frac{1}{4}}{\rho ^{2}}\right] u(\rho )=0
\end{equation}
Note that this expression differs from equation (28) in CC's paper in the
centrifugal term. They obtained $\lambda ^{CC}+1/4$ while here we have $%
\lambda ^{CC}-1/4$. If CC already used their equation (28), then their
results for the Einstein coefficients cannot be correct.

\section{Further comments and conclusions}

\label{sec:comments}

Throughout this paper we solved the eigenvalue equation for an
oversimplified quantum-mechanical model for an electron in the vicinity of
an ionized nanostructure with a permanent electric dipole proposed recently%
\cite{CC19}. Our results suggest that the authors did not solve the angular
part with the intended physical periodic boundary conditions. One may think
that there is just a typo in CC's equation (33) but the fact is that our
results, based on actual functions of period $2\pi $ (present Table~\ref
{tab:lambdas}), do not agree with those in CC's Table~1.

We do not solve the radial part because it is sufficient to show that CC's $%
\lambda $ values are not correct to conclude that their Einstein's
coefficients are surely wrong. However, one can easily show, as we did
above, that the centrifugal term in their equations (28) and (29) should be $%
\frac{\lambda -\frac{1}{4}}{\eta ^{2}}$ instead $\frac{\lambda +\frac{1}{4}}{%
\eta ^{2}}$. We believe that our analysis is correct and that CC's results
may well be meaningless. For example, the Einstein coefficients calculated
by those authors do not correspond with the intended model of $\left(
GaAs\right) _{3}$.

\section*{Acknowledgements}
Paolo Amore acknowledges support from Sistema Nacional de
Investigadores (M\'exico)

\section*{Addendum}

\begin{figure}[tbp]
\begin{center}
\includegraphics[width=9cm]{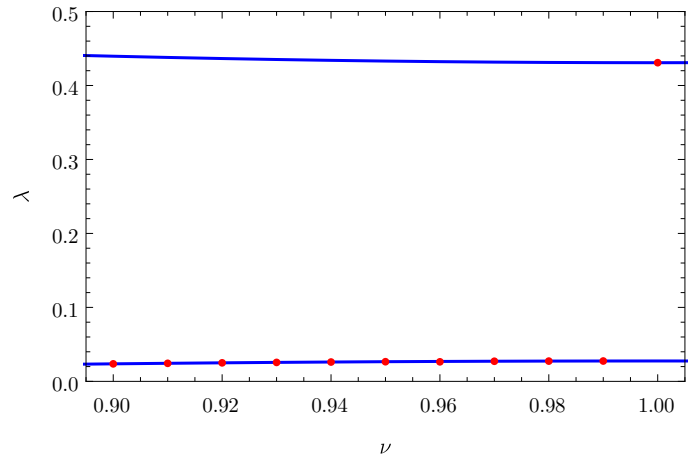} %
\end{center}
\caption{Present results (blue, continuous line) and those of CC\cite{CC20} for $\lambda(\nu)$ and $q=4gp=0.8147872$.}
\label{fig:lambda_vs_nu_comp}
\end{figure}

After the Comment\cite{AF20} and Reply\cite{CC20} were published we could finally reproduce CC's results\cite{CC19,CC20} by simply solving the eigenvalue equation for the Mathieu function as shown by Co\"isson et al\cite{CVY09}. Table~\ref{tab:a_vs_nu} shows $a(\nu)$ for $q=0.8147872$ and $\nu=1,2,3,4$. Boldface entries indicate the values of $a(\nu)$ reported by CC\cite{CC19}. We conjecture that the algorithm used by those authors does not yield the eigenvalues orderly and for that reason they have been picking out the eigenvalues randomly which explains the discontinuity in CC's figure~1\cite{CC19} that does not appear in our more careful calculation given in figure~2 of our comment\cite{AF20}. Figure~\ref{fig:lambda_vs_nu_comp} shows present results $\lambda(\nu)$ (blue, continuous line) and those given by CC\cite{CC20} in their reply (red circles). It is clear that the discontinuity in the figure~1 of their first paper\cite{CC19} is due to a jump from the lowest eigenvalue to the next higher one as conjectured in our Comment\cite{AF20}.

The errors in the calculation of $\lambda$ will obviously affect the results obtained later from the solutions of the radial eigenvalue equation~(\ref{eq:eigen_radial}).

\begin{table}[tbp]
\caption{Eigenvalues of the angular eigenvalue equation (\ref{eq:eigen_rot})
for $\xi=0.8147872/8$}
\label{tab:lambdas}
\begin{center}
\begin{tabular}{cD{.}{.}{10}}

$m$ &    \multicolumn{1}{c}{$2\lambda_m$}       \\
0   &     -0.02038332  \\
1   &     0.9965447876   \\
2   &     1.016922136    \\
3   &     4.001380556    \\
4   &     4.001386528     \\
5   &     9.000592776    \\
6   &     9.000592776    \\

\end{tabular}
\end{center}
\end{table}

\begin{table}[tbp]
\caption{Eigenvalues $a$ of the Mathieu equation for $q = 0.8147872$ and  $\nu=1,2,3,4$}
\label{tab:a_vs_nu}
\begin{center}
\begin{tabular}{cD{.}{.}{10}D{.}{.}{10}D{.}{.}{10}D{.}{.}{10}D{.}{.}{10}}

$\nu$ &    \multicolumn{5}{c}{$a$}       \\
1   &     0.1103083812& \textbf{1}.\textbf{723195887}& 9.033407277& 9.050089309& 25.01383463\\
2   &     -0.3109361980& 3.944835174& \textbf{4}.\textbf{255390446}& 16.02196863& 16.02234952  \\
3   &     0.1103083812& 1.723195887& 9.033407277& \textbf{9}.\textbf{050089309}& 25.01383463  \\
4   &     -0.3109361980& 3.944835174& 4.255390446& 16.02196863& \textbf{16}.\textbf{02234952}  \\

\end{tabular}
\end{center}
\end{table}

\end{document}